\documentclass[12pt,english,journal,onecolumn]{IEEEtran}
\usepackage[T1]{fontenc}
\usepackage[latin9]{inputenc}
\usepackage{geometry}
\usepackage{amsmath}
\usepackage{xpatch}
\usepackage{amssymb}
\usepackage{esint}
\usepackage{mathtools}
\usepackage{amsthm}
\usepackage{nicefrac}
\usepackage{gensymb}
\usepackage{bigints}
\usepackage{physics}
\usepackage{calligra}
\usepackage{graphicx}
\usepackage{dsfont}
\usepackage{array,ragged2e}
\usepackage{enumitem}
\usepackage{etoolbox}
\usepackage{babel}
\usepackage[nopar]{lipsum}
\usepackage{psfrag}
\usepackage[ruled,norelsize]{algorithm2e}
\usepackage{algorithmic}
\usepackage{algorithm2e}
\usepackage{balance}
\usepackage{setspace}
\usepackage{epstopdf}

\SetKw{KwBy}{by}

\makeatletter
\newcommand{\removelatexerror}{\let\@latex@error\@gobble}
\makeatother

\newtheoremstyle{plain}
  {\topsep}   
  {\topsep}   
  {\itshape}  
  {0pt}       
  {\bfseries} 
  {.}         
  {5pt plus 1pt minus 1pt} 
  {\thmname{#1}\thmnumber{ #2} \textnormal{(\thmnote{#3})}} 

\xpatchcmd{\proof}{\hskip\labelsep}{\hskip5\labelsep}{}{}  
\makeatletter
\xpatchcmd{\proof}{\@addpunct{.}}{\@addpunct{:}}{}{}
\makeatother

\newcommand{\noun}[1]{\textsc{#1}}

\renewcommand\[{\begin{equation}}
\renewcommand\]{\end{equation}} 
\usepackage{listings}
\usepackage{fancyvrb}
\usepackage{framed}

\usepackage{courier}
\usepackage[usenames,dvipsnames,table]{xcolor}

\definecolor{dkgreen}{rgb}{0,0.3,0}
\definecolor{gray}{rgb}{0.5,0.5,0.5}

\makeatletter
\patchcmd{\@maketitle}
 {\addvspace{0.5\baselineskip}\egroup}
 {\addvspace{-1.5\baselineskip}\egroup}
 {}
{}

\DeclarePairedDelimiter\floor{\lfloor}{\rfloor}

\begin{document}
\title{\huge Classification Algorithms for Semi-Blind Uplink/Downlink Decoupling in Sub-6 GHz/mmWave 5G Networks}
\author{Hatim Chergui,~\IEEEmembership{Member,~IEEE}, Kamel Tourki,~\IEEEmembership{Senior Member,~IEEE},\\ Redouane Lguensat,~\IEEEmembership{Member,~IEEE},Mustapha Benjillali,~\IEEEmembership{Senior Member,~IEEE},\\ Christos Verikoukis,~\IEEEmembership{Senior Member,~IEEE} and M\'erouane Debbah,~\IEEEmembership{Fellow,~IEEE}
\IEEEcompsocitemizethanks{\IEEEcompsocthanksitem H. Chergui and M. Benjillali are with the Communication Systems Department, INPT, Rabat, Morocco. [e-mail: \{chergui, benjillali\}@ieee.org].\IEEEcompsocthanksitem K. Tourki and M. Debbah are with Huawei Tehnologies Paris Research center, France. [e-mail: \{kamel.tourki, merouane.debbah\}@huawei.com].
\IEEEcompsocthanksitem R. Lguensat is with IGE, Universit\'e Grenoble Alpes, Grenoble, France. [e-mail: redouane.lguensat@univ-grenoble-alpes.fr].
\IEEEcompsocthanksitem C. Verikoukis is with CTTC, Barcelona, Spain. [e-mail: cveri@cttc.es].
}%
}
\maketitle
\thispagestyle{empty}
\begin{abstract}
Reliability and latency challenges in future mixed sub-6 GHz/millimeter wave (mmWave) fifth generation (5G) cell-free massive multiple-input multiple-output (MIMO) networks is to guarantee a fast radio resource management in both uplink (UL) and downlink (DL), while tackling the corresponding propagation imbalance that may arise in blockage situations.
In this context, we introduce a semi-blind UL/DL decoupling concept where, after its initial activation, the central processing unit (CPU) gathers measurements of the Rician $K$-factor---reflecting the line-of-sight (LOS) condition of the user equipment (UE)---as well as the DL reference signal receive power (RSRP) for both 2.6 GHz and 28 GHz frequency bands, and then train a non-linear support vector machine (SVM) algorithm. The CPU finally stops the measurements of mmWave definitely, and apply the trained SVM algorithm on the 2.6 GHz data to blindly predict the target frequencies and access points (APs) that can be independently used for the UL and DL. The accuracy score of the proposed classifier reaches $95\%$ for few training samples.
\end{abstract}

\newpage
\section{Introduction}
\IEEEPARstart{W}{hile} millimeter wave (mmWave) frequency bands are strong candidates for the deployment of 5G massive MIMO wireless systems \cite{r1}, it is also expected, for reliability considerations, that sub-6 GHz band be maintained as a fallback band. Indeed, mmWave offers a large bandwidth to drastically increase the data throughput. However, the mmWave channel propagation presents poor characteristics that weaken the system performance. Furthermore, while the operators can easily deploy high gain antennas \cite{r2}, boost the transmit power, and activate beamforming on access points (APs) to extend their downlink (DL) coverage, the user equipments (UEs) must limit their transmit power and efficiently manage their radio frequency (RF) chains to preserve battery life. Moreover, recent studies \cite{exposure} show that to be compliant with the limits of body exposure to electromagnetic field, the uplink (UL) transmit power at mmWave should be several dB below the power levels in current cellular systems. These limitations threaten to handicap mmWave coverage in UL and would require that operators build ultra-dense networks wherein a high line of sight (LOS) probability is ensured \cite{XLOS} while adopting advanced decoupling techniques that make UL and DL dynamically operating in either sub-6 GHz or mmWave bands using co-located or different APs. In that case, noticeable performance gains might be obtained by associating the UL to a LOS AP where, thanks to the minimal path-loss, the UE can decrease its transmit power allowing the reduction of the UL signal to interference plus noise ratio (SINR) variance, which translates into more efficient and effective UL schedulers and performance gains \cite{DuDe_Chapter}. Nonetheless, Since the UL and DL coverages are quite different, basing the UL and DL associations on the same criterion, which is the DL RSRP, is highly suboptimal \cite{DuDe_Chapter}.
\begin{figure}
\centering
\includegraphics[scale=0.65]{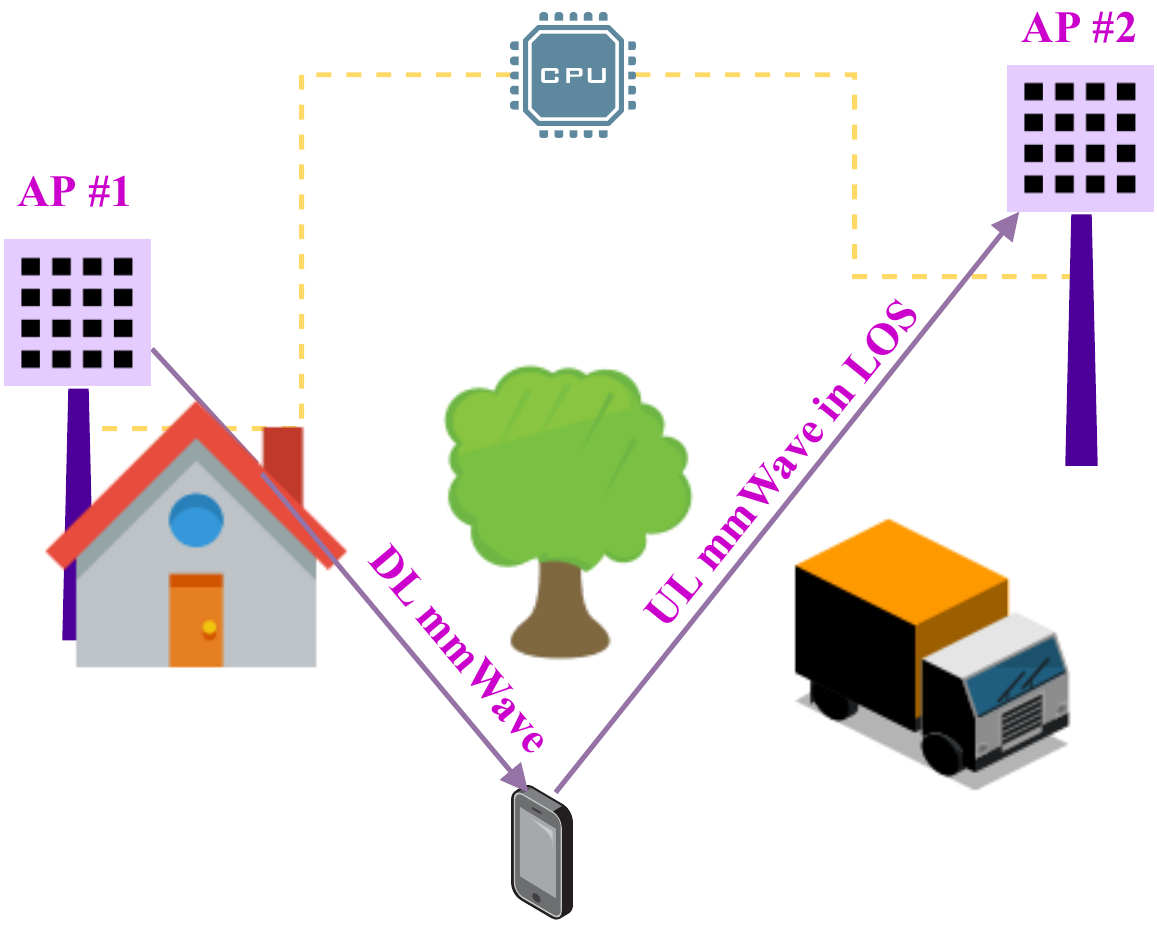}
\caption{The UL uses mmWave with an AP in LOS conditions.}
\vspace{-0.5cm}
\end{figure}

Having said that, frequent measurement reports of both bands are communicated to assist the central processing unit (CPU) decision at the cost of an extra latency as well as a \textit{measurement gap}, where a fraction of data radio resources of the sub-6 GHz band is muted during the measurement of the mmWave one. Therefore, implementing fast machine learning-driven semi-blind decisions in the network turns out to be necessary to sidestep the measurements overhead and latency. 

In \cite{CSIless}, a system referred to as R2-F2 has been introduced, enabling LTE base stations (BSs) to infer the DL channels to a UE by observing the corresponding UL channels. Moerover, the authors in \cite{Integrated} provided a tutorial on integrated mmWave-microwave ($\mu\mathrm{W}$) communications, where they have stated that it is desirable to seamlessly integrate the reliability of $\mu\mathrm{W}$ networks with the high capacity of mmWave networks to achieve ultra-reliable low latency communications (URLLC). In this context, they have shown that reinforcement learning (RL) techniques can be used to exploit information from previous transmissions and assess the reliability of communications over the mmWave band, without frequently performing direct channel measurements. In a self-organizing network (SON) framework, a deep RL model have been proposed in \cite{Reinforce} to improve the DL SINR for an indoor single cell BS operating in mmwave band through iterative actions to increase the SINR from a starting value to a feasible target value. Furthermore, the authors in \cite{Mismar} presented a binary classification algorithm fed by RSRP measurements to predict handover decisions from Sub-6 GHz band to mmWave band yielding a success rate of $91\%$. In \cite{Proactive}, machine learning tools have been leveraged to let the BSs learning how to predict that a certain link will experience blockage in the next few frame periods and proactively hand-over the user to another BS with a highly probable LOS link. In a massive MIMO setup, the authors in \cite{CSI} described how a sample spatial covariance matrix computed from the CSI can be used as an input to a learning algorithm that attempts to relate it to user location.\\

In this letter we investigate the following aspects:
\begin{itemize}
\item First, we incorporate the Rician $K$-factor as an accurate LOS detection metric in 5G measurement reports (MRs). This metric is suitable for UL association as highlighted above.

\item We show that the UL/DL decoupling in a mixed sub-6 GHz/mmWave cell-free antennas cluster can be viewed as a supervised multi-class classification, wherefore a CPU use an already trained and stored non-linear support vector machine (SVM) algorithm to perform decoupling class prediction based on 2.6 GHz measurements only. Subsequently, we refer to it as a semi-blind  UL/DL decoupling algorithm

\item We compare the proposed SVM algorithm performance with its classical threshold-based blind decoupling counterpart, which is similar to the proposed blind handovers in fourth generation (4G) systems.
\end{itemize}
\begin{figure}[t]
\centering
\includegraphics[scale=0.5,trim={0 1cm 0 15.5cm},clip]{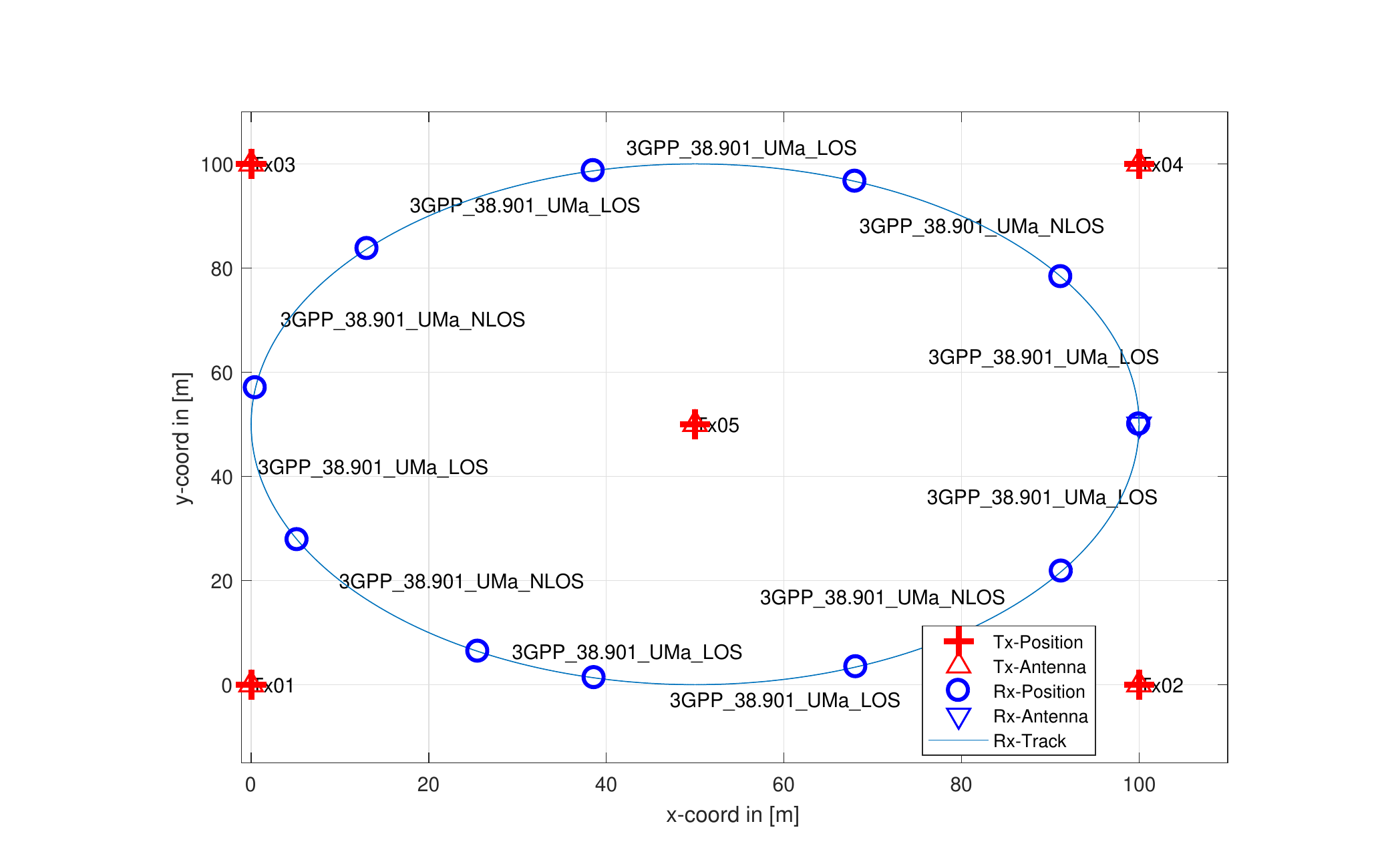}
\caption{Cluster setup and an example of UE track.}
\vspace{-0.5cm}
\end{figure}
\begin{figure}[t]
\centering
\includegraphics[scale=1]{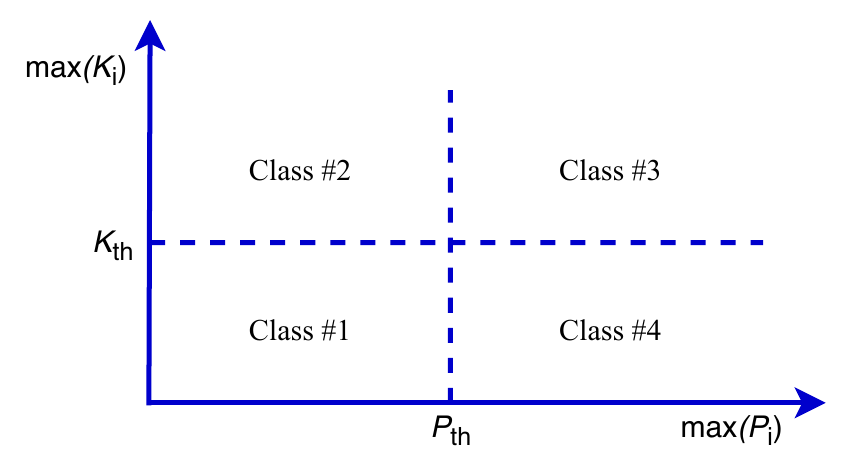}
\caption{Data (mmWave band) labeling using $K_{\mathrm{th}} =3\,\mathrm{dB}$ and $P_{\mathrm{th}} =-115\,\mathrm{dBm}$.}
\vspace{-0.2cm}
\end{figure}
\section{Dataset Establishment}
Using the well-established QuaDRiGa simulator \cite{Quadriga}, we generate a dataset corresponding to a mixed sub-6 GHz/mmWave cluster, where 5 APs are all equipped with 8 and 64-antenna arrays operating at 2.6 GHz and 28 GHz, respectively, and placed at a height of 25 m as depicted in Fig. 2. Their radiation patterns follow the 3GPP models \cite{TR36873} and \cite{TR38901}, respectively. The UE is endowed with an omni-directional antenna and follows either a linear or circular track with alternating LOS and non-LOS (NLOS) conditions, modeled according to the 3GPP urban macro (UMa) LOS/NLOS framework \cite{TR38901} with a pedestrian velocity of 1 m/s. 

During initial CPU activation, it collects the measurement reports of both 2.6 GHz and 28 GHz, containing the RSRPs and $K$-factors with respect to the five APs. The measurement data is then labeled based on the mmWave $K$-factors and RSRPs, and we define four classes, depicted in Fig. 3, as follows:

\begin{itemize}
\item \textbf{Class 1:} UL and DL at 2.6 GHz.
\item \textbf{Class 2:} UL at 28 GHz with a LOS AP, and DL at 2.6 GHz with the best serving AP.
\item \textbf{Class 3:} UL 28 GHz with a LOS AP, and DL at 28 GHz with the best serving AP.
\item \textbf{Class 4:} UL at 2.6 and DL at 28 GHz.
\end{itemize}

The CPU then stops the measurement of the mmwave band definitely, and takes semi-blind decoupling decisions based on 2.6 GHz MRs only, where the corresponding 5 $K$-factors and 5 RSRPs are the 10 \emph{features} to be used by the machine learning algorithm as detailed in the sequel. The UE finally selects the best target APs lying in the decision class sent back by the CPU.

\section{Semi-Blind Decoupling Algorithm}
\subsection{Exploratory Data Visualization and Algorithm Choice}
To unveil the nature of the boundaries that might separate the samples, we proceed with a data visualization. Since the dataset contains 10 features, we invoke the principal component analysis (PCA) approach \cite{PCA}, where we project the data into a 3-dimensional (3D) space via a linear transformation. Given that the samples, depicted in Fig. 4, can be separated only using non-linear surfaces, it follows that the prospective machine learning algorithm should be a non linear classifier. Also, to enable a fast decision, we assume that the CPU should settle for a small training dataset of 2.6 GHz/28 GHz measurements sent by different UEs simultaneously. As such, SVMs with non linear radial basis function (RBF) kernel turns out to be a suitable scheme \cite{SVM}, since only support vectors (not all the samples) are used to construct the hyperplanes separating the samples.
\begin{figure}[h!]
\centering
\includegraphics[scale=0.7,trim={0 5cm 0 10cm},clip]{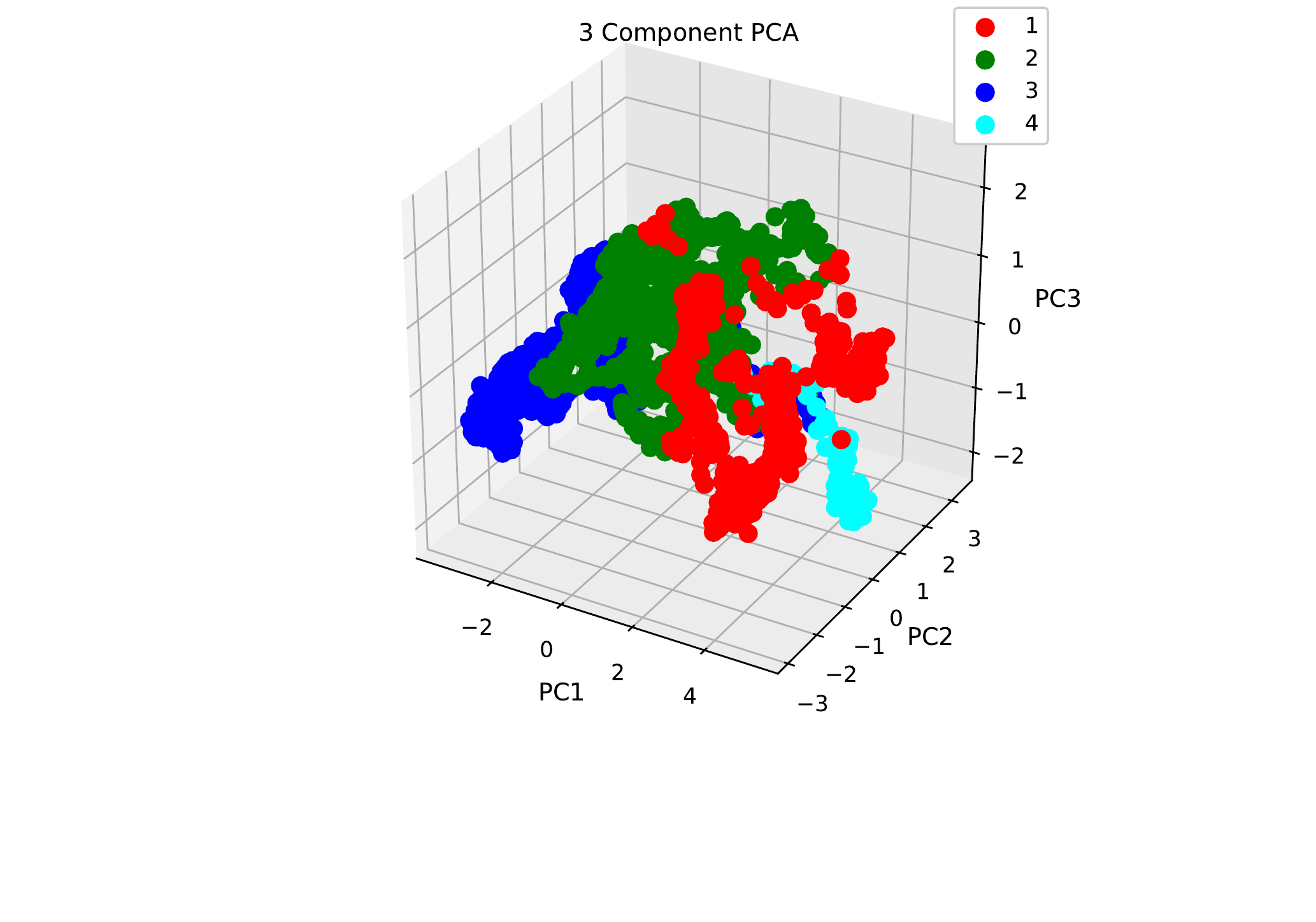}
\vspace{-1.3cm}
\caption{PCA of the dataset. The classes are non-linearly separable.}
\end{figure}

\begin{figure}[h!]
\centering
\includegraphics[scale=1,trim={0 14.5cm 0 11cm},clip]{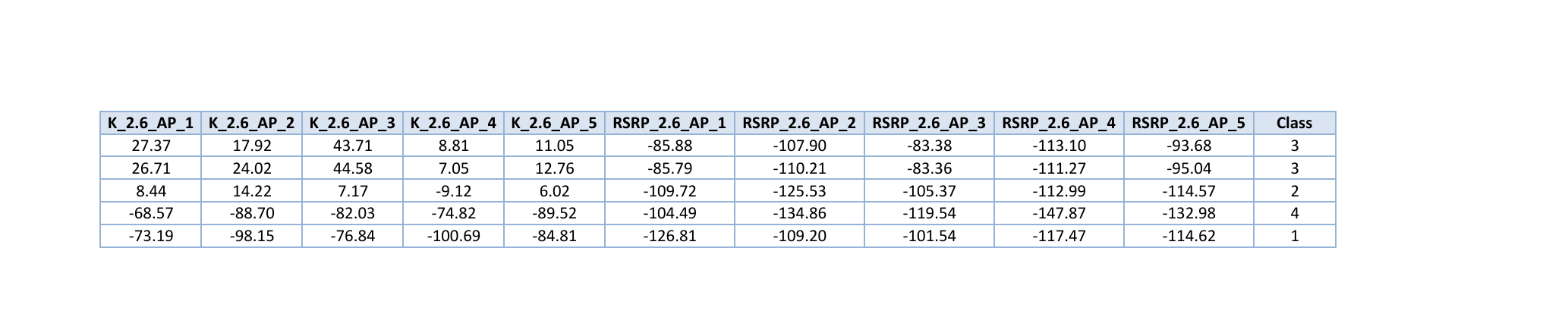}
\caption{An extract from the dataset used to train the SVM classifier during the initial CPU activation.}
\end{figure}
\subsection{Support Vector Machine}
The proposed machine learning algorithm is detailed in Algorithm 1. It has been implemented using the widespread package Scikit-learn \cite{Scikit}. During the training phase, the CPU shuffles the collected data to ensure an even distribution of the class labels and quickly reach a high accuracy using only few training samples. To achieve this goal, the built-in \texttt{shuffle} method from \texttt{numpy} has been used. 

It is worth mentioning that the decoupling success rate has been calculated across different training set sizes to highlight the algorithm fast convergence. The parameters of the adopted non-linear RBF-based SVM are listed in Table I. The \texttt{C} parameter trades off misclassification of training examples against the simplicity of the decision surface. A low \texttt{C} makes the decision surface smoother, while a high \texttt{C} aims at classifying all training examples correctly by giving the model freedom to select more samples as support vectors at the expense of an over-fitting. The \texttt{gamma} parameter is the inverse of the learning rate, representing the influence of a single training sample on the model.

\begin{algorithm}
  \caption{SVM Decoupling Success Rate (DSR)}
  \textbf{Inputs:} $N=3800$ samples dataset, consisting of $K$-factors and RSRPs at 2.6 GHz and the corresponding classes as depicted in Fig. 5.\\
  \textbf{Outputs:} The decoupling success rate.
 
  \begin{algorithmic}[1]
   \STATE Select the first $N_T=3000$ samples for training and shuffle them
   \STATE Separate the training features $X_{T}$ from their class numbers $y_{T}$ 
   \STATE $\mathrm{DSR}\gets \left[\,\right]$
   \FOR{$i\gets50$ \KwTo $N_T$ \KwBy $50$}
         \STATE Standardize $X_{T}^{(i)}$
         \STATE Initialize SVM with RBF kernel and train it using $X_{T}^{(i)},y_{T}^{(i)}$
         \STATE $L\gets$ Number of data samples per decision window
         \STATE $N_{\mathrm{test}}\gets\floor*{(N-N_{T})/L}$
         \STATE $\alpha\gets0$
         \FOR{$j := 1$ to $N_{\mathrm{test}}$}
            \STATE Extract the test data $X_{t}^{(j)},y_{t}^{(j)}$
         	\STATE Predict $\hat{y}_{t}^{(j)}$ using $X_{t}^{(j)}$
         	\STATE Calculate the Accuracy Score ($\mathrm{AS}$), where $\mathrm{AS}\left(y_t^{(j)},\hat{y}_{t}^{(j)}\right)=\frac{1}{L}\sum_{l=0}^{L-1} \mathds{1}\left(y_{t,l}^{(j)}=\hat{y}_{t,l}^{(j)}\right)$
         	 \IF{$\mathrm{AS}>0.9$}
      			\STATE Consider $\hat{y}_{t}^{(j)}$ a valid decision
      			\STATE $\alpha\gets \alpha + 1$
      		 \ENDIF
        \ENDFOR
      \STATE $\alpha_\mathrm{avg}\gets \alpha/N_{\mathrm{test}}$  
      \STATE $\mathrm{DSR}\gets \left[\mathrm{DSR}\,\, \alpha_\mathrm{avg}\right]$
   \ENDFOR
  \end{algorithmic}
\end{algorithm}

\begin{table}[!htb]
\vspace{-4mm}
\label{Table1}
\centering	
\newcolumntype{M}[1]{>{\centering\arraybackslash}m{#1}}

\caption{SVM Parameters}
\begin{tabular}{m{5cm}M{4.5cm}}
\hline 
\multicolumn{1}{>{\centering\arraybackslash}M{5cm}}{\cellcolor{black!20} Parameter} & \multicolumn{1}{>{\centering\arraybackslash}M{4.5cm}}{\cellcolor{black!20} Value}\\
\hline 
\texttt{kernel} & \texttt{rbf}\\
$\mathtt{C}$ & $\mathtt{1}$\\
\texttt{gamma} & \texttt{0.4}\\
\texttt{decision\_function\_shape} & \texttt{ovo}\\

\hline
\hline
\end{tabular}
\end{table}
\begin{figure}[h!]
\centering
\includegraphics[scale=0.62,trim={0 9cm 0 10cm},clip]{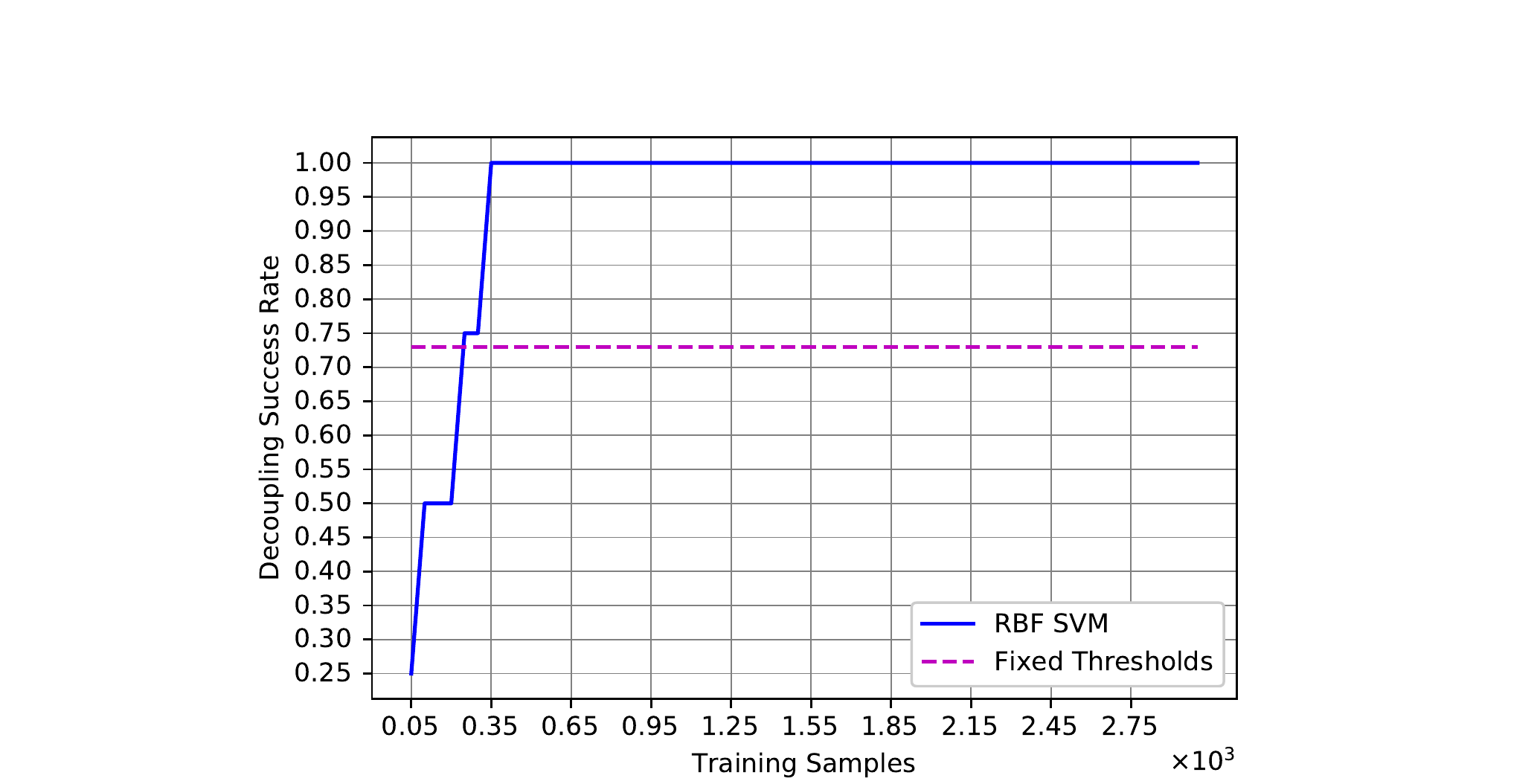}
\caption{Decoupling success rate vs. the number of training samples.}
\end{figure}
\section{Performance Assessment}
In this section, we present the performance assessment of the proposed algorithm compared to a benchmark.\\
Fig. 6 depicts the decoupling success rate of the presented SVM-based strategy. A comparison to the classical threshold-based scheme (benchmark) have been detailed in Algorithm 2. In this case, it is worth noting that the classical blind decoupling is triggered once the source frequency $K$-factor or RSRP varies with respect to its preset threshold. Similarly to blind handovers in 4G systems, the offsets are added to these fixed thresholds to account for the difference between the source and the target bands. Let $K_1,\ldots,K_5$ and $P_1,\ldots,P_5$ denote the measured metrics at 2.6 GHz for the 5 APs, respectively. A blind decoupling decision might be triggered, for instance, if
\begin{equation}
\max(K_i) > K_\mathrm{th} + K_\mathrm{offset}.
\end{equation}

In this context, the dataset statistics show that the 2.6 GHz $K$-factor is lower than the 28 GHz one with an average offset of $-5.03 \,\mathrm{dB}$. However, the RSRP of the former is higher than the latter's one with an average offset of $23.24 \,\mathrm{dB}$. This could be justified by the use of narrowbeam massive antenna arrays at 28 GHz and the low path-loss at 2.6 GHz.

Note that since the decoupling success rate is defined as the probability of reaching an accuracy score of $90\%$, we get a step-wise variation in Fig.~6.

On the other hand, we compare in Fig.~7 the accuracy score of the RBF and linear kernel SVM algorithms. We notice the convergence of the former to $95\%$ using only 350 samples for training, which guarantees a fast and accurate decoupling decision. Furthermore, the linear-kernel SVM fails to converge and its accuracy score is largely underperforming the RBF-based SVM, which confirms the algorithm choice we have made in Section III.
\begin{figure}[h!]
\centering
\includegraphics[scale=0.62,trim={0 9cm 0 10cm},clip]{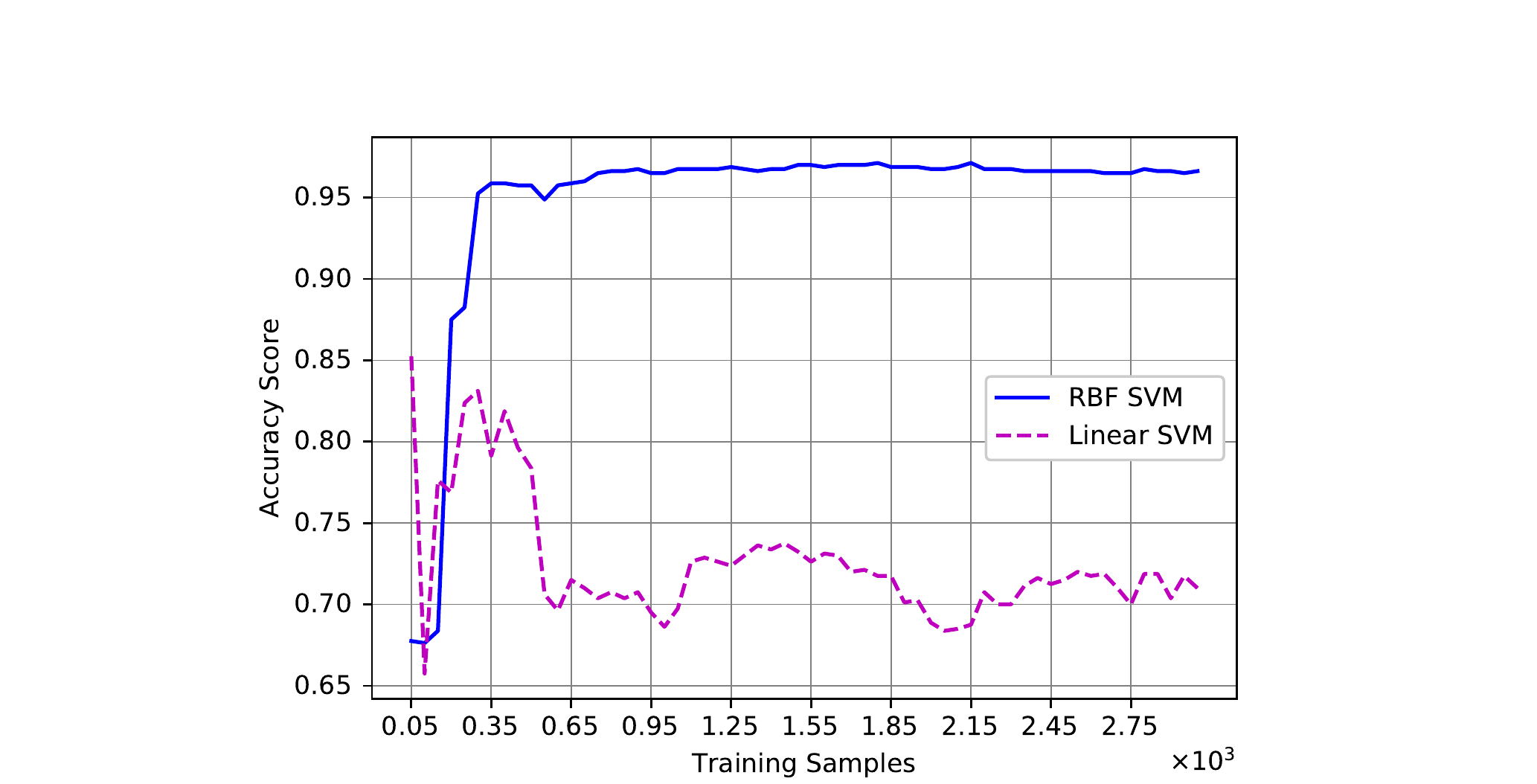}
\caption{Accuracy score vs. the number of training samples.}
\end{figure}
\begin{algorithm}
  \caption{Fixed Thresholds DSR}
  \textbf{Inputs:} $N$ samples dataset, consisting of $K$-factors and RSRPs at both 2.6 GHz and 28 GHz.\\
  \textbf{Outputs:} The decoupling success rate.
 
  \begin{algorithmic}[1]
   \STATE Label the dataset by applying the criteria of Fig.~3 \\on 28 GHz data.
   \STATE Estimate the mean offsets of $K$-factors and RSRPs between 2.6 GHz and 28 GHz over the dataset.
   \STATE To infer the decoupling class from 2.6 GHz data, use the same criteria of Fig.~3 while replacing $P_\mathrm{th}$ with $P_\mathrm{th}+P_\mathrm{offset}$ and $K_\mathrm{th}$ with $K_\mathrm{th}+K_\mathrm{offset}$.
   \STATE Compare the obtained classes with the dataset \\28 GHz labels and calculate the average decoupling accuracy.
  \end{algorithmic}
\end{algorithm}


\section{Conclusion}
In this letter, we have presented a machine learning algorithm to enable fast semi-blind UL and DL decoupling in mixed sub-6 GHz/mmWave cell-free 5G networks. We have shown that choosing a decoupling scheme blindly translates into a non-linear classification problem, wherefore we have adopted RBF-kernel support vector machine to solve it. The performance thereof have been compared with its classical threshold-based counterpart which is similar to 4G blind handovers. The proposed scheme reaches $95\%$ of accuracy for only few training samples, which makes it a fast and reliable approach to automate UL and DL decoupling in 5G networks.
\appendix{\noun{~~~~~~~~~~~~~~~~~~~~~~~~~~~~~~~~~~~~~~~~~~~~~~~~~~~~~~~~~~~~Dataset and Source Code}}

The full Python source code as well as the dataset are available in \cite{Code}.

\balance

\end{document}